\documentclass[conference]{IEEEtran}
\IEEEoverridecommandlockouts

\usepackage{color}
\usepackage{bm} 
\usepackage{amsmath,amsfonts}
\usepackage{algorithmic}
\usepackage{algorithm}
\usepackage{array}
\usepackage{textcomp}
\usepackage{stfloats}
\usepackage{url}
\usepackage{cite}
\usepackage{verbatim}
\usepackage{graphicx}
\usepackage{balance}
\usepackage{ulem}
\usepackage{float} 
\usepackage{subfigure}
\usepackage{setspace}
\usepackage{amssymb}
\usepackage{siunitx}
\usepackage[numbers,sort&compress]{natbib}
\usepackage{hyperref}
\def\BibTeX{{\rm B\kern-.05em{\sc i\kern-.025em b}\kern-.08em
    T\kern-.1667em\lower.7ex\hbox{E}\kern-.125emX}}
\begin{document}
\graphicspath{{Figs/}}

\title{Sensing Performance Analysis in Cooperative Air-Ground ISAC Networks for LAE
}

\author{\IEEEauthorblockN{Yihang Jiang\IEEEauthorrefmark{1}\IEEEauthorrefmark{2},
Xiaoyang Li\IEEEauthorrefmark{2},
Guangxu Zhu\IEEEauthorrefmark{1}, 
Xiaowen Cao\IEEEauthorrefmark{3},
Kaifeng Han\IEEEauthorrefmark{4},
Bingpeng Zhou\IEEEauthorrefmark{5},
Xinyi Wang\IEEEauthorrefmark{6}
}
\IEEEauthorblockA{\IEEEauthorrefmark{1}Shenzhen Research Institute of Big Data, The Chinese University of Hong Kong (Shenzhen), Guangdong, China}
\IEEEauthorblockA{\IEEEauthorrefmark{2}Southern University of Science and Technology, Guangdong, China}
\IEEEauthorblockA{\IEEEauthorrefmark{3}Shenzhen University, Guangdong, China}
\IEEEauthorblockA{\IEEEauthorrefmark{4}China Academy of Information and Communication Technology, Beijing, China}
\IEEEauthorblockA{\IEEEauthorrefmark{5}Sun Yat-sen University (Shenzhen), Guangdong, China}
\IEEEauthorblockA{\IEEEauthorrefmark{6}Beijing Institute of Technology, Beijing, China}
}

%


\maketitle

\begin{abstract}
To support the development of low altitude economy, the air-ground \textit{integrated sensing and communication} (ISAC) networks need to be constructed to provide reliable and robust communication and sensing services. In this paper, the sensing capabilities in the cooperative air-ground ISAC networks are evaluated in terms of area radar detection coverage probability under a constant false alarm rate, where the distribution of aggregated sensing interferences is analyzed as a key intermediate result. Compared with the analysis based on the strongest interferer approximation, taking the aggregated sensing interference into consideration is better suited for pico-cell scenarios with high base station density. Simulations are conducted to validate the analysis.
\end{abstract}

\begin{IEEEkeywords}
Integrated sensing and communication, low altitude economy, network modeling, performance analysis, stochastic geometry.
\end{IEEEkeywords}

\section{Introduction}
The \textit{low-altitude economy} (LAE) has emerged as a promising concept that utilizes the near-ground airspace to facilitate innovative aerial applications with significant socioeconomic potential \cite{jiang2025integrated}. Ensuring the sustainable development of this emerging sector demands robust safety assurance frameworks, underpinned by pervasive communication infrastructure and real-time sensing capabilities. These technological requirements create a natural synergy of \textit{integrated sensing and communication} (ISAC) systems, which is recognized as a critical enabler for LAE applications \cite{fei2023air}.

As a fundamental technology for LAE development, ISAC achieves deep integration of traditionally separated radar sensing and wireless communication functions within a unified system architecture. This innovative technological convergence delivers substantial performance improvements through shared utilization of spectrum resources, hardware platforms, and signal processing capabilities \cite{liu2022integrated}. Meanwhile, the system modeling and performance analysis of network-level ISAC provides critical insights for practical deployment of multi-cell wireless networks, enabling optimized configurations to meet the diverse operational requirements of LAE scenarios.

Recent studies have systematically investigated various aspects of ISAC network performance under different operational conditions. The impact of interference on ISAC systems was quantitatively examined in \cite{ren2018performance}, employing a time-division approach that allocates distinct temporal intervals for \textit{communication and sensing} (C\&S) functions, respectively. Building upon this foundation, \cite{sun2024performance} extended the analysis to incorporate the effects of urban building obstructions on network performance metrics. A comprehensive analytical framework was developed in \cite{olson2023coverage} to evaluate both C\&S capabilities, deriving closed-form expressions for joint coverage probability and ergodic rate in ISAC-enabled networks. 

In above existing studies focusing on terrestrial networks, the critical impacts of altitude on C\&S were neglected. Also, the radar detection coverage probability that satisfies a certain \textit{constant false alarm rate} (CFAR) level have not been studied. To address this research gap, our recent work \cite{jiang2024coverage, jiang2025network} has pioneered the investigation of integrated air-ground ISAC networks. In \cite{jiang2025network}, a comprehensive analytical framework for network-level performance evaluation in cooperative air-ground ISAC systems has been proposed. With this framework, the C\&S performances were evaluated in terms of different metrics including area communication coverage probability, area communication spectral efficiency, \textit{area radar detection coverage probability} (ARDCP) under a CFAR criterion, and average Cram\'er-Rao bound. Although our previous study in \cite{jiang2025network} adopted the \textit{strongest interferer approximation} (SIA) method to evaluate the ARDCP under a CFAR criterion, and demonstrated its close alignment with simulation results in low \textit{base station} (BS) density scenarios, a thorough performance analysis under more general conditions remains an open problem. To address this limitation, this paper provides a comprehensive analysis of the aggregated sensing interference characterization for the aforementioned cooperative air-ground ISAC networks. 

\textit{Notations:} Boldface lowercase and uppercase letters denote vectors and matrices, respectively. For any \textit{random variable} (RV) $X$, we employ $\mathbb{E}\{X^n\}$ to denote its $n$-order moment, $\kappa_{X}(n)$ for the $n$-order cumulant, and $\mathbb{P} \{X >T \}$ for the probability of event $X >T$, respectively. Specific distributions are indicated by: $X\sim \text{Exp}(\lambda)$ (exponential), $X\sim \mathcal{S} (\alpha, \beta, c, \mu)$ (stable) and $X\sim \mathcal{I}\mathcal{G} (\rho , \varrho )$ (inverse Gaussian). Standard mathematical functions include $\mathbf{1}(\cdot)$ (indicative), $\text{exp}(\cdot)$ (exponential), $\text{sgn}(\cdot)$ (signum), $\text{sinc}(\cdot)$ (normalized sinc), $\Gamma(\cdot)$ (Gamma), and $\text{erfc}(\cdot)$ (complementary error).

\section{Networks Framework}
\begin{figure}[!t]
  \centering
  \includegraphics[scale = 0.4]{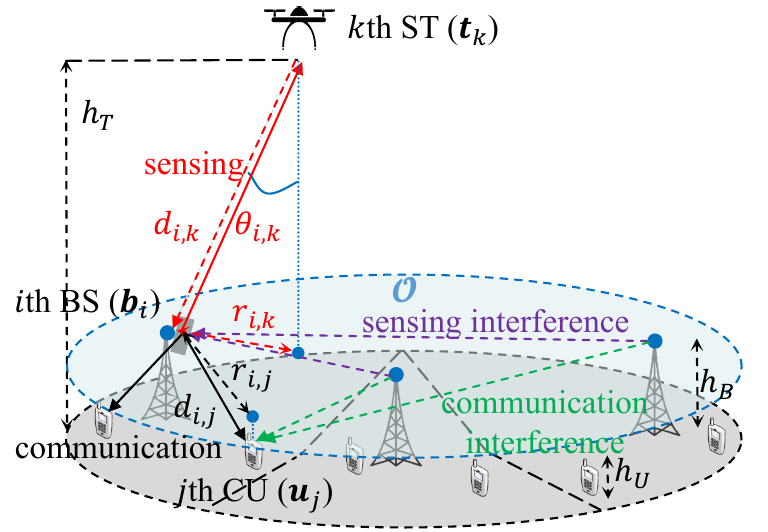}
  \caption{Air-ground ISAC networks with terrestrial CUs and aerial STs \cite{jiang2025network}.}
  \label{FigSys}
  \vspace{-0.5cm}
\end{figure}

\subsection{System Model}
As shown in Fig.~\ref{FigSys}, we consider a downlink air-ground ISAC network composed of BSs, terrestrial \textit{communication users} (CUs), and aerial \textit{sensing targets} (STs). Each BS is equipped with a vertical uniform linear array of $N_t$ transmit antennas for dual-functional signal transmission and $N_r$ receive antennas for radar echo reception. The ISAC signals serve both single-antenna CUs and sense non-cooperative aerial STs simultaneously. BSs are spatially distributed according to a 2-dimensional (2D) homogeneous \textit{Poisson point process} (PPP) $\Phi_{B}$ with density $\lambda_B \ \mathrm{BSs/Km}^2$, while CUs and STs follow mutually independent stationary point processes. The heights of the BSs, CUs, and STs are denoted as $h_B \ \mathrm{m}$, $h_U \ \mathrm{m}$ and $h_T \ \mathrm{m}$, respectively, with $h_T>h_B>h_U $. Each BS serves multiple CUs and senses multiple STs under a nearest-neighbor association policy.

With the stochastic geometry framework, we analyze the average network performance by focusing on a typical BS $i$, CU $j$ and ST $k$ \cite{andrews2011tractable}. Using the horizontal plane $\mathcal{O}$ where BSs are located as a reference, the geometric model is illustrated in Fig. \ref{FigSys}. The positions of the BS, CU, and ST are denoted as $\bm b_i = (x_i, y_i, h_B)$, $\bm u_j = (x_j, y_j, h_U)$, and $\bm t_k = (x_k, y_k, h_T)$, respectively. Let $r_{i, j}^{(c)}$ and $r_{i, k}^{(r)}$ denote the horizontal distances from BS $i$ to CU $j$ and ST $k$, respectively. The vertical separations are $\Delta h_c = h_B - h_U$ and $\Delta h_r = h_T - h_B$. The corresponding 3D distances are $d_{i,j}^{(c)} = \sqrt{\left(r_{i,j}^{(c)}\right) ^2 + \Delta h_c^2 }$ and $d_{i,k}^{(r)} = \sqrt{\left(r_{i,k}^{(r)}\right) ^2 + \Delta h_r^2 }$. For simplicity, we use $r_i$ and $d_i$ to denote the horizontal and 3D distances from BS $i$ to the typical node (CU or ST). The angle of departure (for sensing) or arrival (for echo) is given by $\theta_{i,k} = \text{arcos} \frac{\Delta h_r}{\left\lVert \bm b_i - \bm t_k\right\rVert_2 } = \text{arcsin} \frac{r_{i,k}}{\left\lVert \bm b_i - \bm t_k\right\rVert_2 }$.  
\subsection{Signal Model}
The system operates at carrier frequency $f_c$ over a total bandwidth $B$, divided into $N$ OFDM subcarriers with spacing $\Delta f$. Each OFDM symbol has duration $T = T_g + T_s$, where $T_g$ is the guard interval and $T_s = 1/\Delta f$ is the useful symbol time. Over a \textit{coherent processing interval} (CPI) consisting of $M\times N$ time-frequency resource blocks, each BS serves $J$ CUs and senses $K$ STs within its cell using the same ISAC signals. The channel is assumed stationary during the CPI. 

\subsubsection{Communication Model}
Each CU is randomly assigned $N/J$ subcarriers. Following the literature \cite{olson2023coverage}, we analyze the performance defined on an arbitrary resource element in the context of communication. The serving BS is labeled as BS 1, and it allocates equal power to all its served CUs. The received signal at a typical CU is 
\begin{equation}
  \begin{aligned}
    \label{comm signal}
  y_c &= \sqrt{P_t \mathcal{L}_1^c}\bm h^H_1 \bm x_1 
   + \underbrace{\sum_{i \in \Phi_{B} | 1 } \sqrt{P_t\mathcal{L}_i^c} \bm h^H_i \bm x_i }_{\text{inter-cell interference} } + z_1 ,
\end{aligned}
\end{equation}
where $P_t$ is the transmit power, $\mathcal{L}_i^c = \left(\frac{\lambda_c}{4\pi}\right)^2 d_i^{- \alpha_c } $ is the communication path loss with exponent $\alpha_c$ and wavelength $\lambda_c$, $\bm h_i^H \in \mathbb{C}^{1\times {N_t} } $ is the communication channel vector with every entry following $ \mathcal{C} \mathcal{N} (0, 1)$, $\bm x_1 = \bm w_1 s_1$ is the precoded signal with unit-power symbol $s_1$ and \textit{beamforming} (BF) vector $\bm w_1 \in \mathbb{C}^{{N_t}\times 1}$, and $z_1\sim \mathcal{C}\mathcal{N}(0, \sigma_c^2 )$ is the additive white Gaussian noise with variance $\sigma_c^2$. 

\subsubsection{Sensing Model}
Sensing utilizes the entire time-frequency resources within a CPI. We consider monostatic sensing in LoS conditions. The parameters of interest are relative distance $d_{rel} = c\tau/2$ and relative speed $v_{rel} = \lambda_c f_D/2$, where $\tau$ is time delay and $f_D$ is Doppler frequency. We assume $f_c\gg B$, $f_D \gg \Delta f$ and $\tau < T_g$ to avoid inter-symbol interference. Using the periodogram method~\cite{sturm2011waveform}, parameter estimation is transformed into a spectral analysis problem. We focus on a typical ST. The radar observation on the $(m,n)$-th resource element is
\begin{align}
    \label{raw radar observation}
 &\left[\bm Y_r\right]_{m,n} \!=\! \sqrt{P_t \mathcal{L}_1^r} \bm f_1^H(\theta_1) \underbrace{\bm a_r(\theta_1)\bm a_t^T(\theta_1)}_{\triangleq \bm G_1(\theta_1) } \bm w_1 \left[\bm S_1\right]_{m,n}\! e^{\jmath 2\pi mT f_D } \nonumber \\ 
 & \cdot e^{-\jmath 2\pi n \Delta f\tau } \!\!\! + \!\!\!\! \underbrace{ \sum_{i \in \Phi_{B} | 1 } \!\!\! \sqrt{P_t \mathcal{L}_{1,i}^c} \underbrace{\bm f_1^H(\theta_1) \bm H _{1,i}}_{\triangleq \bm \nu_{1,i} ^H(\theta_1)} \bm w_{i} \left[\bm S_i\right]_{m,n} \! e^{-\jmath 2\pi n \Delta f\tau_{1,i} } }_{\text{inter-BS interference} } \nonumber \\
 &+ \left[\bm Z_1\right]_{m,n},
\end{align}
where $\mathcal{L}_1^r = \frac{\lambda_c^2\xi }{(4\pi)^3} d_1^{-2\alpha _r} $ is the sensing path loss with radar cross-section $\xi$ and exponent $\alpha_r$, $\bm G_1(\theta_1) = \bm a_r(\theta_1) \bm a_r^T(\theta_1)$ is the steering matrix with receive steering vector $\bm a_r(\theta_1) = [1, \cdots, e^{\jmath \pi (N_r-1)\cos \theta_1}]^T \in \mathbb{C} ^{N_r\times 1}$ and transmit steering vector $\bm a_t(\theta_1) = [1, \cdots, e^{\jmath \pi (N_t-1)\cos \theta_1}]^T \in \mathbb{C} ^{N_t \times 1}$,  $\mathcal{L}_{1,i}^c$ denotes the path loss between the sensing interference BS $i$ and the BS 1, $\bm \nu_{1,i} ^H(\theta_1) = \bm f_1^H(\theta_1) \bm H _{1,i} \in \mathbb{C} ^{1 \times N_t}$ is the combined interference channel with each entry in $\bm H _{1,i} \in \mathbb{C} ^{N_r \times N_t}$ distributed as $\mathcal{C}\mathcal{N}(0, 1)$, and $\left[\bm Z_1\right]_{m,n} \sim \mathcal{C}\mathcal{N}( 0, \sigma_r^2 )$ is the Gaussian noise. 

To decouple the symbol effect, we apply a phase rotation:
\begin{align}
  \label{sensing signal}
  &\left[\widetilde{\bm Y} _r\right]_{m,n} = \left[\bm Y_r\right]_{m,n} / \left[\bm S_1\right]_{m,n} \\
  &= \sqrt{P_t \mathcal{L}_1^r} \bm f_1^H(\theta_1) \bm G_1(\theta_1)\bm w_1 e^{\jmath 2\pi mT f_D }e^{-\jmath 2\pi n \Delta f\tau } + \nonumber \\ 
  & \sum_{i \in \Phi_{B} | 1 } \sqrt{P_t \mathcal{L}_{1,i}^c } \bm \nu _{1,i}^H(\theta_1) \bm w_{i} \left[\widetilde{\bm S_i} \right]_{m,n} e^{-\jmath 2\pi n \Delta f\tau_{1,i} }  
  + \left[\widetilde{\bm Z} _1\right]_{m,n} \nonumber ,
\end{align}
where $\widetilde{\bm S_i}$ and $\widetilde{\bm Z} _1$ are the phase-adjusted symbol and noise.

\subsection{Cooperative Beamforming Design}
Sensing echo signals experience round-trip path loss, making them significantly weaker than the direct interference signals transmitted by nearby BSs. To mitigate this severe sensing interference, a cooperative BF strategy is designed. Specifically, a cluster of $N_c + 1$ nearest BSs centered at BS 1, denoted as $\Phi_{CB}$, is formed fto collaboratively design BF vectors. At the receiver side of each BS $i\in\left\{1, \cdots, N_c+ 1\right\} $, maximum ratio combining is applied using the receive BF vector
\begin{equation}
  \bm f_i^H(\theta_{i,k}) = \frac{1}{\sqrt{N_r}}[1, \cdots, e^{-\jmath \pi (N_r-1)\cos \theta_{i,k}} ]\in \mathbb{C} ^{1\times N_r} ,
  \end{equation}
where $\theta_{i,k}$ is the angle between BS $i$ and its ST $k$. Since a linear combination of complex Gaussian RVs remains complex Gaussian, the combined sensing interference channel $\bm \nu _{i,1}^H(\theta_{i,k})$ for all $ i \in \left\{2, \cdots, N_c + 1\right\} $ follows a Rayleigh distribution. Leveraging this property, cooperative \textit{zero-forcing} (ZF) BF is implemented at BS 1 to nullify the sensing interference toward the other $N_c$ BS. This is achieved by projecting the BF vector $\bm w_1$ onto the null space of the aggregated combined sensing interference channels.

To suppress interference toward the $N_c$ nearest BSs during sensing respective $K$ STs, BS 1 must sacrifice $KN_c$ degrees of freedom. The aggregated interference channel matrix is constructed as
\begin{align}
  &\bm{G}_I = \left[\left(\bm \nu_{2,1}^H(\theta_{2,1})\right)^T, \left(\bm \nu_{2,1} ^H(\theta_{2,2})\right)^T, \cdots, \left(\bm \nu_{2,1} ^H(\theta_{2,K})\right)^T, \cdots, \right. \nonumber \\
  &\!\!\!\!\!\!\!\! \phantom{=\;\;}\left. \left(\bm \nu_{{N_c+1},1}^H(\theta_{{N_c+1},1})\right)^T \!\!, \cdots \!, \left(\bm \nu_{{N_c+1},1}^H(\theta_{{N_c+1},K})\right)^T\right] \! \in \! \mathbb{C} ^{N_t\times KN_c} \!.
\end{align} 
The normalized cooperative ZF BF matrix at BS 1 is then given by
\begin{equation}
 \bm{W}_1 = \frac{\bm{\widetilde{H}}_1^H \left(\bm{\widetilde{H}}_1 \bm{\widetilde{H}}_1^H \right)^{-1} }{ \left\lVert\bm{\widetilde{H}}_1^H \left(\bm{\widetilde{H}}_1 \bm{\widetilde{H}}_1^H \right)^{-1} \right\rVert_2 } ,
\end{equation} 
where $\bm{\widetilde{H}}_1 \triangleq  \left[\left(\bm h^H_1\right)^T, \bm{G}_I\right]^T $. The BF vector $\bm w_1$ is extracted from the first column of $\bm{W}_1$.

Under this cooperative framework, the horizontal distance from the central BS 1 to any interfered BS outside the cluster satisfies $r_I \geq r_{N_c + 2}$, where $r_{N_c + 2}$ is the horizontal distance from BS 1 to the $(N_c + 2)$-th nearest BS. This condition in turn ensures that a lower bound on the interference distance between interfering BS and BS 1.

\section{Sensing Performance Evaluation}
In this paper, we focus on the sensing performance analysis in the interference-limited networks, where noise is ignored due to severe interference in dense cell scenarios. Thus, according to \eqref{sensing signal}, the radar sensing \textit{signal-to-interference ratio} (SIR) for the typical ST can be denoted as 
\begin{equation}
\begin{aligned}
  \label{sensing SIR}
   \gamma _{r} = \frac{\mathcal{L}_1^r N_r \underbrace{\left\lvert \bm a_t^T(\theta_1) \bm w_1 \right\rvert^2}_{\triangleq g_{rs}} }{ \sum\limits_{i \in \Phi_I \triangleq \Phi_{B} | \Phi_{CB} } \mathcal{L}_{1,i}^c \underbrace{\left\lvert \bm \nu _{1,i}^H(\theta_1) \bm w_{i} \right\rvert^2}_{ \triangleq g_{ri} } } 
  = \frac{\xi (4\pi)^{-1} N_r d_1^{-2\alpha _r} g_{rs} }{ \underbrace{\sum_{i \in \Phi_I } w_i^{-\alpha _c} g_{ri}}_{\triangleq I_{ar}} } ,
\end{aligned}
\end{equation} 
where $\Phi_{I}$ is the PPP set of sensing interfering BSs, $g_{rs} $ and $g_{ri} $ represent the BF gain for the sensing and the interference channels, respectively, $d_1$ is the link distance from BS 1 to the typical ST, $w_i$ is the distance from sensing interference BS $i$ to BS 1, and $I_{ar}$ denotes the aggregated sensing interference. Based on \cite{jiang2025network}, the beamforming gains are modeled as $g_{rs} \sim \text{Exp}(1)$ and $g_{ri} \sim \text{Exp}(1)$. 

\subsection{Sensing Performance Metric}

Mathematically, for each bin in the periodogram, the false alarm happens if $P_t\frac{\lambda_c^2}{(4\pi)^2} I_{ar} > \eta'$ with $\eta'$ denoting the threshold. Therefore, the CFAR for each bin is defined as
\begin{equation}\label{Pbin}
  \mathcal{P}_{CFAR, bin} \triangleq  \mathbb{P} \left\{P_t \frac{\lambda_c^2}{(4\pi)^2}I_{ar} > \eta'\right\}  = \mathbb{P} \{ I_{ar} > \eta \}.
\end{equation}
The corresponding CFAR during the CPI of an OFDM frame can be expressed as
\begin{equation}\label{Pframe}
  \mathcal{P}_{CFAR, frame} =  1 - \left(1 - \mathcal{P}_{CFAR, bin}\right)^{NM}.
\end{equation}
Note that the definition of $\mathcal{P}_{CFAR, frame}$ is based on the duration of one frame following \cite{Braun14}. The frame-based performance metric is also adopted in \cite{olson2023coverage}.

Given the required CFAR $\mathcal{P}_{CFAR, frame}$, the threshold $\eta$ can be determined from \eqref{Pbin} and \eqref{Pframe} based on the distribution of the aggregated interference $I_{ar}$. The corresponding required SIR can be further obtained as $T_r = \eta/\mathbb{E} \left\{I_{ar}\right\} $.\footnote{The SIR threshold $T_r$ is a statistic, in which it is assumed that $\mathbb{E} \left\{I_{ar}\right\}$ is a constant over sensing procedure and can be obtained by pre-measurement.} Therefore, the corresponding ARDCP under given CFAR can be defined as
\begin{equation}
  \mathcal{P}_{arcov} \triangleq \lambda_B K\mathbb{P} \left\{NM\gamma _{r}>T_r \right\}  ,
\end{equation}
which represents the density of detected STs that satisfies certain CFAR level, where $NM$ comes from the signal processing gain of periodogram \cite{sturm2011waveform}.  

\subsection{Aggregated Sensing Interference Distributions}
Before going into the detailed analysis, we first characterize the distribution of the aggregated sensing interference. This will be a key intermediate result in the ARDCP analysis. To address this problem, our previous work \cite[Proposition 3]{jiang2025network} applies the SIA to obtain an analytic expression between $\mathcal{P}_{CFAR,frame}$ and $T_r$. The approximation that only considers the strongest interference source, while ignoring interference from other BSs, becomes less accurate when the BS density is high.


According to \eqref{sensing SIR}, the aggregated sensing interference in cooperative scenarios is defined as  
\begin{equation}
  \begin{aligned} 
    \label{truncated sensing interference}
  I_c \triangleq I_{ar} = \sum\nolimits_{i \in \Phi_I } w_i^{-\alpha _c} g_{ri}  .
\end{aligned}
\end{equation} 

Directly characterizing the distribution of $I_c$ is analytically intractable due to its complex form. To facilitate analysis, we first examine a non-cooperative scenario in which interferers are uniformly distributed across the Euclidean plane, i.e.,
\begin{equation}
  \begin{aligned} 
    \label{total sensing interference}
I_{nc} \triangleq \sum\nolimits_{i \in \Phi _B | 1 }  w_i^{-\alpha _c}g_{ri} .
\end{aligned}
\end{equation} 
This simplified model enables a tractable derivation of the interference distribution in the non-cooperative case. By comparing \eqref{total sensing interference} and \eqref{truncated sensing interference}, it can be observed that $I_c$ corresponds to a truncated version of $I_{nc}$ with a guard interval defined by $r_c \triangleq r_{N_c + 2}$. Building on this insight, we first analyze the non-cooperative interference and then use the results to approximate the distribution of the cooperative interference, which is the main focus of this paper. Here, $I_c$ and $I_{nc}$ denote the truncated aggregated sensing interference in cooperative scenarios and the total aggregated sensing interference in non-cooperative scenarios, respectively.

\textbf{\textit{1) Approximated Distribution of $I_{nc}$:}}

\textit{\textbf{Lemma 1} (Stable distribution \cite{haenggi2012stochastic}):} 
A stable distribution $X = \mathcal{S}(\alpha, \beta, c, \mu )$ has a \textit{characteristic function} (CF) of the following form
\begin{equation}
  \begin{aligned}
    \label{Stable distribution}
    \varphi _X(\omega) = \text{exp}\left(\jmath \omega\mu - c \left\lvert \omega\right\rvert^{\alpha} \left[1 - \jmath\beta \text{sgn}\ \omega \Phi (\alpha , \omega) \right]\right), 
  \end{aligned}
\end{equation}
where $\alpha $ denotes the stability parameter for characterizing the tail thickness, $\beta $ denotes the skewness parameter for characterizing the distribution symmetry, $c$ denotes the scale parameter, $\mu$ denotes the location parameter, $\text{sgn}(\cdot)$ denotes the sign function, and 
\begin{equation}
  \begin{aligned}
\Phi(\alpha , \omega) = \left\{
  \begin{array}{rcl}
  \text{tan}\left(\frac{\pi \alpha }{2}\right)  ,& & \alpha \neq 1 \\
  - \frac{2}{\pi}\log \left\lvert \omega\right\rvert  , & & \alpha = 1
  \end{array} \right.
\end{aligned}
\end{equation}
denotes the phase correction function. 

Since the potential sensing interferers are randomly and uniformly distributed throughout the entire Euclidean plane, the interference distance can be infinite, hence we follow a basic idea from \cite{haenggi2012stochastic} to derive the distribution of aggregated sensing interference in non-cooperative case. The corresponding CF is given in the following proposition.

\textit{\textbf{Proposition 1} (Zero-centered right-skewed stable distribution of $I_{nc}$):}
The CF of $I_{nc}$ can be derived as
\begin{equation}
  \begin{aligned}
    I_{nc} = \mathcal{S}\left(\alpha = \frac{2}{\alpha_c} , \beta = 1 , c = \frac{\lambda_B \pi}{\text{sinc}\ \alpha _c^{-1}}, \mu = 0 \right)  ,
  \end{aligned}
\end{equation}
which corresponds to a zero-centered right-skewed stable distribution. The specific details is given in the following proof.

\textit{Proof:} Our proof proceeds in two key stages. First, we examine a finite network confined within a disk region $\mathcal{D} (\bm O, r_f) \triangleq \mathcal{A}_{r_f}$ of radius $r_f$ centered at the location of the sensing echo receiving BS $\bm O$, which is conditioned on a fixed number of interferers with independently and identically distributed locations. Subsequently, we remove this cardinality condition and take the limit as the disk radius $r_f \rightarrow \infty$ to obtain the complete interference characterization.

\textbf{\textit{Stage 1: Finite Network Analysis}}. We first consider interference from sources located within $\mathcal{A}_{r_f}$:
\begin{equation}
  \begin{aligned}
    I_{r_f} = \sum\nolimits_{i \in \mathcal{A}_{r_f} } w_i^{-\alpha _c} g_{ri} \triangleq \sum\nolimits_{i \in \mathcal{A}_{r_f} } \ell (w_i),
  \end{aligned}
\end{equation} 
where the distribution of $w_i$ follows the \textit{probability density function} (pdf): 
\begin{equation}
  \begin{aligned}
    \label{sensing interference pdf}
    f_{W}(w) = \frac{2w }{r_f^2} \mathbf{1}\left(0\leq w \leq r_f\right) 
  \end{aligned} .
\end{equation}
Note that $I_{r_f} \rightarrow I_{nc}$ when $r_f\rightarrow \infty$. 

The CF of $I_{r_f}$ can be defined as 
\begin{equation}
  \begin{aligned}
\varphi _{I_{r_f}}(\omega) \triangleq \mathbb{E}_{I_{r_f}} \left\{\text{exp}\left(\jmath\omega I_{r_f}\right) \right\}  .
  \end{aligned}
\end{equation}

Conditioning on $p$ interferers within $\mathcal{A}_{r_f}$, we obtain
\begin{equation}
  \varphi _{I_{r_f}}(\omega) = \mathbb{E} \left\{\mathbb{E}_{I_{r_f}} \left\{\text{exp}\left(\jmath\omega I_{r_f}\right) | \mathcal{N}(\mathcal{A} _{r_f}) = p \right\}\right\} .  
\end{equation}
Given $p$ as uniformly distributed points in $\mathcal{A}_{r_f}$ with the radial density defined in \eqref{sensing interference pdf}, the CF becomes the product of $p$ individual CFs as
\begin{equation}
  \begin{aligned}
    \label{conditional Expect}
  \mathbb{E}_{I_{r_f}} \!\!\left\{e^{\jmath\omega I_{r_f}} | \mathcal{N}(\mathcal{A} _{r_f}) \!=\! p \right\} = \left(\int_{0}^{r_f} \frac{2w}{r_f^2 } \mathbb{E}_{g_{ri} } \left\{e^{\jmath\omega \ell (w)}\right\} \,dw \right) ^p \!\!.
\end{aligned}
\end{equation}

Since the number of nodes in $\mathcal{A} _{r_f}$ follows a Poisson distribution, we have
\begin{equation}
  \begin{aligned}
    \label{finite CF}
  &\varphi _{I_{r_f}}(\omega) = \sum\nolimits_{p = 0}^{\infty} \underbrace{\frac{\text{exp}\left(-\lambda_B \pi r_f^2\right) \left[\lambda_B \pi r_f^2\right] ^p }{p!}}_{\text{the probability of $p$ points exist within region $\mathcal{A} _{r_f}$ }} \\
  &\cdot \mathbb{E}_{I_{r_f}} \left\{\text{exp}\left(\jmath\omega I_{r_f}\right) | \mathcal{N}(\mathcal{A} _{r_f}) = p \right\} .
\end{aligned}
\end{equation}

Substituting \eqref{conditional Expect} into \eqref{finite CF}, and recognizing the Taylor expansion of the exponential function, one can get
\begin{equation}
  \begin{aligned}
  \varphi _{I_{r_f}}(\omega) = \text{exp}\left(\lambda_B \pi r_f^2 \left[ \int_0^{r_f} \frac{2w}{r_f^2 } \mathbb{E}_{g_{ri} } \left\{e^{\jmath\omega \ell (w)}\right\} \,dw  - 1 \right] \right) .
\end{aligned}
\end{equation}

\textbf{\textit{Stage 2: Infinite Network Analysis}}. Taking the limit as $r_f\rightarrow \infty$ and conditioning on $g_{ri}$, one can apply integration by parts with the substitution $w = \ell^{-1}(x)$ to obtain
\begin{equation}
  \begin{aligned}
&\lim_{r_f \to \infty} r_f^2 \left[ \int_0^{r_f} \frac{2w}{r_f^2 } e^{\jmath\omega \ell (w)} \,dw - 1 \right]  \\
&= \int_{0}^{\infty } \left(\ell^{-1}(x) \right)^2 \jmath\omega \ e^{\jmath\omega x} \,dx .  
\end{aligned}
\end{equation}

This yields the conditional CF as
\begin{equation}
  \begin{aligned}
    \label{conditional CF}
    &\varphi _{I_{nc} | g_{ri} }(\omega) = \text{exp}\left(\lambda_B \pi \int_{0}^{\infty} \left(\ell^{-1}(x) \right)^2 \jmath\omega \ e^{\jmath\omega x} \,dx \right)  \\
    &= \text{exp}\left(\lambda_B \pi g_{ri}^{\frac{2}{\alpha _c}} \jmath\omega \int_{0}^{\infty } x^{-\frac{2}{\alpha _c}} e^{\jmath\omega x} \,dx\right) \\
    &\overset{\text{(a)}}{= } \text{exp}\left(-\lambda_B \pi g_{ri}^{\frac{2}{\alpha _c}} \Gamma \left(1 - \frac{2}{\alpha _c} \right) \left(-\jmath\omega \right)^{\frac{2}{\alpha _c}}\right)  , 
  \end{aligned}
\end{equation}
where (a) follows from the property of the Gamma function under the condition that $\alpha _c > 2$ and $\omega \geq 0$, with the result for negative $\omega$ obtained via the symmetry condition $\varphi _{I_{nc}}(\omega) = \varphi _{I_{nc}}^*(-\omega)$. 

Taking expectation over $g_{ri} $, we obtain
\begin{equation}
  \begin{aligned}
    &\varphi _{I_{nc}}(\omega) = \text{exp}\left(-\lambda_B \pi \mathbb{E} _{g_{ri}}\left\{g_{ri}^{\frac{2}{\alpha _c}}\right\}  \Gamma \left(1 - \frac{2}{\alpha _c} \right) \left(-\jmath\omega \right)^{\frac{2}{\alpha _c}}\right) \\
    &\overset{\text{(a)}}{= } \text{exp}\left(- \frac{\lambda_B \pi}{\text{sinc}\ \alpha _c^{-1}} \left\lvert \omega\right\rvert ^{\frac{2}{\alpha _c}} \left[1 - \jmath\text{sgn}\ \omega\tan\frac{\pi}{\alpha _c} \right] \right) , 
  \end{aligned}
\end{equation}
where (a) comes from the fact that $\mathbb{E}\left\{x^a\right\} = \Gamma(1 + a)/\lambda^a$ for exponential $x$ with parameter $\lambda$. The CF corresponds to a skewed stable distribution with parameters $\mu = 0$, $c = \lambda_B \pi/\text{sinc}\ \alpha _c^{-1}$, $\alpha = 2/\alpha_c$, and $\beta = 1$. Since $\mu = 0$ and $\beta > 0$, this distribution can be recognized as zero-centered and right-skewed.
$\hfill \blacksquare$

\textbf{\textit{2) Approximated Distribution of $I_{c}$:}} \\
By comparing $I_c$ and $I_{nc}$, we observe that the integral counterpart in \eqref{conditional CF} is capped at $\ell(r_c)$ since the interferences within $\mathcal{A}_{r_f}$ are canceled. Hence, $I_c$ can be effectively modeled by a so-called \textit{truncated-stable distribution} (TSD) \cite{rosinski2007tempering}
\begin{equation}
  \begin{aligned}
    \label{TSD}
    \varphi _{I_c}(\omega) = \text{exp}\left(c_{I_c} \Gamma(-\alpha_{I_c} )\left[(g_{I_c} - \jmath \omega)^{\alpha_{I_c}} - g_{I_c}^{\alpha_{I_c}} \right]\right)   , 
  \end{aligned}
\end{equation} 
where $c_{I_c}$ and $\alpha_{I_c}$ are mirror stable distribution parameters, $g_{I_c}$ tempers the tail decay. 

The final step involves determining the parameters of the TSD. Notably, as the cooperative radius $r_c \to 0$, $I_c\to I_{nc}$, which follows a stable distribution with characteristic exponent $\alpha = 2/\alpha_c$. This asymptotic behavior motivates our choice to fix the TSD's characteristic exponent as $\alpha_{I_c} = 2/\alpha_c$. With $\alpha_{I_c}$ established, the remaining parameters can be derived through cumulants matching, as presented in the following proposition.

\textit{\textbf{Proposition 2} (Cumulants Matching of TSD):} 
By matching the first two cumulants of the TSD with those of the aggregated sensing interference, the distribution parameters (given $\alpha_{I_c}$) are determined as
\begin{subequations}
  \label{parameters of TSD}
  \begin{align}
    c_{I_c} &= \frac{- \kappa _{I_c} (1)}{\Gamma(-\alpha_{I_c}) \alpha_{I_c} \left[\frac{\kappa _{I_c} (1)(1 - \alpha_{I_c})}{\kappa _{I_c} (2)} \right]^{\alpha_{I_c} - 1} }  ,  \\
    g_{I_c} &= \frac{\kappa _{I_c} (1)(1 - \alpha_{I_c}) }{\kappa _{I_c} (2)} .
  \end{align}
\end{subequations}

\textit{Proof:}
The $n$-th cumulant of the TSD follows from \eqref{TSD} as
\begin{equation}
  \begin{aligned}
    \label{n-order cumulant of TSD}
        \kappa (n) &= \jmath^{-n} \frac{d^ n}{d \omega^n} \ln \varphi _{I_c}(\omega) \bigg |_{\omega = 0} \\
        &= (-1)^n c_{I_c} \Gamma(-\alpha_{I_c} ) g_{I_c}^{\alpha_{I_c} - n} \prod\nolimits_{i = 0}^{n - 1} (\alpha_{I_c} - i)  .
  \end{aligned}
\end{equation}

Through Campbell's theorem \cite{haenggi2012stochastic}, the $n$-th cumulant of the truncated aggregated interference can be expressed as
\begin{equation}
  \begin{aligned}
    \label{n-order cumulant of interference}
    \kappa_{I_c}(n) = \frac{2\pi \lambda_B}{n\alpha_c - 2}r_c^{2 - n\alpha_c} \mathbb{E} \{g_{ri}^n \} = \frac{2\pi \lambda_Br_c^{2 - n\alpha_c} \Gamma(1+n)}{n\alpha_c - 2} .
  \end{aligned}
\end{equation}
 
Equating these cumulant expressions yields the parameter solutions for $c_{I_c}$ and $g_{I_c}$, which are given in terms of the first two cumulants of $I_c$ as \eqref{parameters of TSD}.
$\hfill \blacksquare$

\textit{\textbf{Remark 1}:} SIA is more suitable for scenarios characterized by low BS density. In such scenarios, the distribution of BSs is sparse, which means that interference from the strongest interferer (typically the one closest to the receiver) will dominate the aggregated interference. Consequently, the aggregated interference distribution exhibits high kurtosis, reflecting its peakedness and concentration around the mean. Additionally, due to the influence of distant interference sources, the tail of the distribution demonstrates a heavy-tailed characteristic with slow decay. In contrast, the TSD approximation is more suitable for scenarios with high BS density. Here, BSs are clustered together, each serving a smaller coverage area. As a result, the interference from different sources are relatively uniform, leading to an aggregated interference distribution with low kurtosis. This distribution appears more uniform and symmetric, with thinner tails compared to the SIA case. 

\subsection{Sensing Performance Analysis}
Once the required sensing SIR threshold $T_r$ is determined, the corresponding ARDCP under CFAR is given in the following theorem.

\textit{\textbf{Theorem 1 (ARDCP under CFAR in Air-Ground Cooperative ISAC Network)}}: The ARDCP under CFAR criterion of the air-ground cooperative ISAC network is expressed as
\begin{equation}
  \begin{aligned}
\mathcal P_{arcov}(T_r) = \lambda_B K \int_{0}^{\infty } \mathcal{P}_{rcov | R_{1}}(T_r, r_1) f_{R_{1}}( r_1) \,d r_1  ,
\end{aligned}
\end{equation}
where $\mathcal{P}_{rcov | R_{1}}(T_r, r_1)$ denotes the link-level conditional coverage probability for radar sensing, which is given by
\begin{equation}
  \begin{aligned}
    &\mathcal{P}_{rcov | R_{1}}(T_r, r_1) \triangleq  \mathbb{P}\left\{ NM\gamma _{r} > T_r | R_{1} =  r_1\right\} \\
    &= \frac{2\pi \overline{r}_{N_c + 2}^{2-\alpha _c}}{\alpha _c - 2} \ {_2F_1} \left(1, 1 - \frac{2}{\alpha _c}; 2 - \frac{2}{\alpha _c}; - q \overline{r}_{N_c + 2}^{-\alpha _c}\right),
  \end{aligned}
\end{equation} 
where $q = \frac{4\pi T_r d_1^{2\alpha _r}}{NMN_r\xi}$, $_2F_1(a , b; c; z ) = F(a , b; c; z )$ denotes the Gauss hypergeometric function \cite{gradshteyn2014table}, $\overline{r}_{N_c + 2}$ denotes the expectation of $r_{N_c + 2}$, and $f_{R_{1}}(r_1) = 2\pi\lambda_B r_1 e^{-\pi \lambda_B r_1^2 } $ denotes the pdf of horizontal distance from BS 1 to the typical ST. 

\textit{Proof:} See \cite[Appendix E]{jiang2025network}.  $\hfill \blacksquare$

\section{Simulation Results}
In this section, numerical simulations are conducted to characterize the sensing interference distribution and analyze the ARDCP in cooperative air-ground ISAC networks. The theoretical framework presented earlier is validated through extensive \textit{Monte Carlo} (MC) simulations. Unless otherwise specified, all simulation parameters align with the reference configuration provided in \cite[Table I]{jiang2025network}.

\begin{figure}[!t]
  \vspace{-0.3cm}
  \centering
  \subfigure[Case 1: SIA under low BS density scenarios]{
  \includegraphics[scale=0.33]{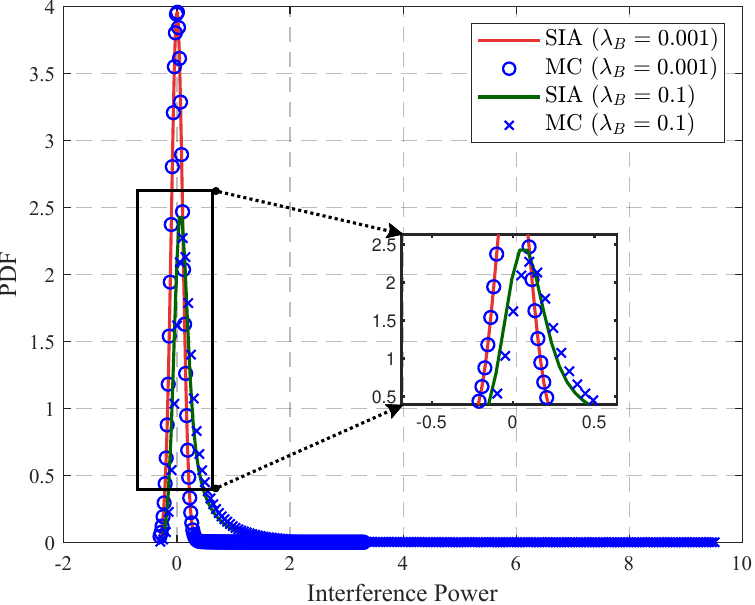}}
  \subfigure[Case 2: TSD approximation under high BS density scenarios]{
  \includegraphics[scale=0.33]{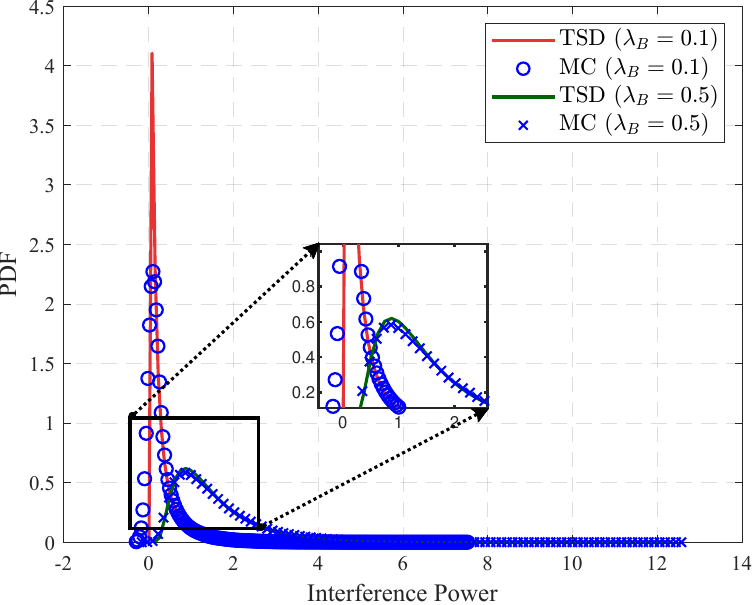}}
  \caption{Distribution approximation of the aggregated sensing interference under different cases, where $\lambda_B$ is in units of $\mathrm{BSs/m}^2$.}
  \vspace{-0.2cm}
  \label{interference distribution}
\end{figure}

The simulation results in Fig.~\ref{interference distribution} validate the accuracy of the proposed approximations, where the SIA closely matches MC simulations in scenarios with low BS density (capturing the high-kurtosis characteristic), while the TSD approximation accurately fits scenarios with high BS density (modeling heavy-tailed distributions), demonstrating their respective suitability for sparse and dense network deployments. 

\begin{figure}[!t]
  \vspace{-0.3cm}
  \centering
  \subfigure[Case 1: SIA under low BS density scenarios]{
  \includegraphics[scale=0.33]{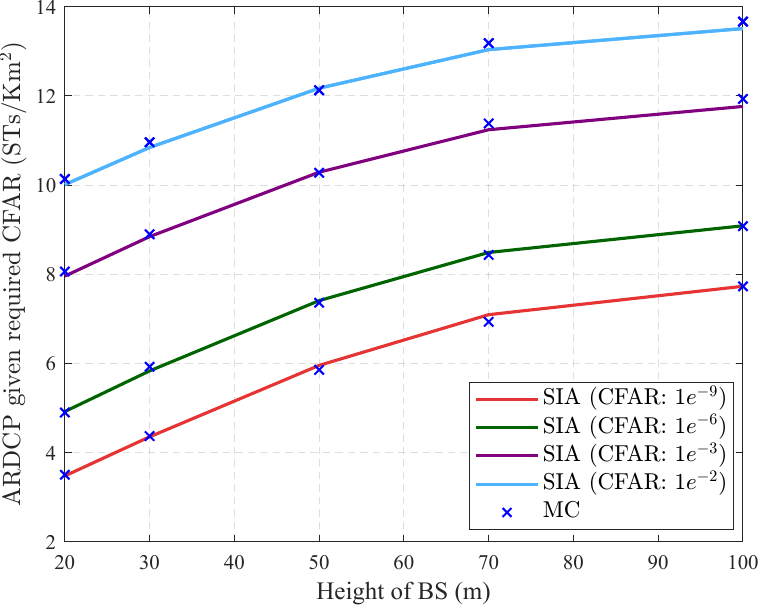}}
  \subfigure[Case 2: TSD approximation under high BS density scenarios]{
  \includegraphics[scale=0.33]{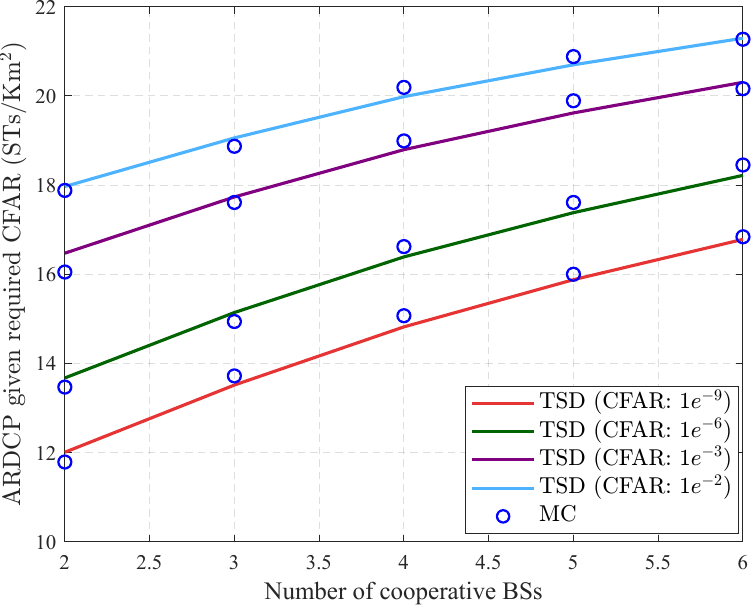}}
  \caption{ARDCP with varying BS heights and cooperative BS cluster sizes.}
  \label{ARDCP}
  \vspace{-0.3cm}
\end{figure}

Fig.~\ref{ARDCP} demonstrates the impact of BS height and cooperative BS cluster size on ARDCP performance under varying CFAR requirements. The consistence of simulation and analytical results validates our theoretical framework. In sparse network scenarios (Case 1), ARDCP improves with increasing BS height due to reduced BS-to-ST distances and consequently higher SIR. For dense deployments (Case 2), where interference dominates system performance, ARDCP enhancement is primarily achieved through expansion of the cooperative BS cluster size.

\section{Conclusion}
In this work, the ARDCP under the CFAR criterion was analyzed as a key sensing metric within a cooperative air-ground ISAC networks. A refined analytical approach was developed to approximate the distribution of the aggregated sensing interference. Furthermore, the properties and applicable conditions of both the SIA and the TSD approximation were examined through numerical simulations. This paper contributes to the promising new area of air-ground cooperative wireless networks for ISAC by analyzing the key sensing indicator, while many interesting follow-up research issues warrant further investigation, such as the scenarios of collaborated sensing and bi-static sensing.

\normalem
\bibliographystyle{IEEEtran}
\bibliography{refs}

\end{document}